\documentstyle[aps,prl]{revtex}

\begin{document}
\draft

\twocolumn[\hsize\textwidth\columnwidth\hsize
\csname@twocolumnfalse\endcsname

\title{Electronic specific heat in the normal state of cuprate high-$T_c$
superconductors}
\author{F. Mancini and D. Villani}
\address{Dipartimento di Scienze Fisiche ''E. R. Caianiello'' - Unit\`a I.N.F.M. di
Salerno\\
Universit\`a di Salerno, 84081 Baronissi (SA), Italy}
\author{H. Matsumoto}
\address{Institute for Material Research, Tohoku University, Sendai 980, Japan}
\date{December 16, 1996}
\maketitle

\begin{abstract}\widetext
By means of the composite operator method we calculate the specific heat of
the 2D Hubbard model. Our results show that the model gives a detailed
description of the experimental situation in the normal state of cuprate
high $T_c$ superconductors. The observed behavior is consistent with a Fermi
liquid picture where the unusual properties are described in terms of the
van Hove scenario. There is experimental evidence, based on the data for the
spin magnetic susceptibility and specific heat, that this scenario is
related to the overdoped region where superconductivity is depressed.
\end{abstract}

]

\narrowtext

A remarkable property of cuprate materials is the presence of a van Hove
singularity (vHs) near the Fermi level in the hole-doped compounds [1].
Based on this experimental result a van Hove scenario for high $T_c$
superconductors has been proposed [2-4]. Now, there are two quantities, the
coefficient of the electronic specific heat $\gamma =C/T$ and the spin
magnetic susceptibility $\chi ({\bf k},\omega )$, which are directly
connected to the density of states (DOS) and can give important information
on the role played by the vHs. In our previous articles [5-7] we have
presented a series of theoretical results where we showed that the
single-band 2D Hubbard model can give a detailed description of the normal
state magnetic properties observed in high $T_c$ materials. In particular,
the picture that emerges from our studies [6,7] of the magnetic
susceptibility is consistent with the van Hove scenario, where the unusual
observed behavior is explained by the fact that upon doping the Fermi level
approaches the vHs.

In this Letter we shall examine the electronic specific heat and we will
show that also in this case the Hubbard model is a realistic model capable
to describe the thermodynamic properties of the oxides in the framework of
the van Hove scenario.

The electronic specific heat $C(T)$ of cuprate high $T_c$ superconductors
has been measured. In particular $C(T)$ of $La_{2-x}Sr_x CuO_4$ [8] has been
studied for $0.03<x<0.44$ in the range of temperatures between $1.5$ and $%
300K$, and of $YBa_2Cu_3O_{6+y}$ [9] for $0.16\le y\le 0.97$ between $1.8$
and $300 K$. In Ref. 10 the electronic specific heat of $La_{2-x}Ba_x CuO_4$
has been measured in the range $0.025<x<0.25$ at low temperatures. From
these experiments the following behavior has been observed for the
coefficient $\gamma=C/T$ of the normal state specific heat:

\begin{description}
\item  {a.} for fixed temperature, $\gamma (x,T)$ increases with doping;

\item  {(a1)} in the case of $La_{2-x}Sr_xCuO_4$, $\gamma (x,T)$ exhibits a
rather sharp maximum at $x\approx 0.25$ (near the doping where
superconductivity disappears), then starts to decrease; for $%
La_{2-x}Ba_xCuO_4$ a maximum has been observed at $x\approx 0.22$;

\item  {(a2)} in the case of $YBa_2Cu_3O_{6+y}$, $\gamma (x,T)$ increases
smoothly to a plateau or two broad maxima, situated at $y\approx 0.6$ and $%
y\approx 0.9$, respectively;

\item  {b.} for fixed doping, $\gamma (x,T)$ as a function of temperature
exhibits a broad peak moving to lower temperatures with increasing the
dopant concentration;

\item  {c.} further increasing $y$, the T dependence weakens and in the
region of high doping no increase is observed. For $YBa_2Cu_3O_{6+y}$, no
substantial increase is observed for $y>0.8$.
\end{description}

To interpret these results, let us recall that the coefficient of the
electronic specific heat $\gamma=C/T$ can be expressed as 
\begin{equation}
\gamma(T)={\frac{1}{T}} {\frac{d}{dT}} \int_{-\infty}^{+\infty} d\omega
N(\omega) f(\omega) \omega
\end{equation}
where $N(\omega)$ is the density of states and $f(\omega)$ is the Fermi
distribution function. This expression explicitly shows that $\gamma(T)$
reflects thermal average of the density of states. The presence of a maximum
(or two maxima) for a critical doping is related to the fact that the Fermi
energy lies on the vHs. Interpretation of the experimental results obtained
in Ref. 9 for $YBa_2Cu_3O_{6+y}$ in terms of a sharp feature in the density
of states, consistent with ARPES experiments [1], was firstly advanced in
Refs. 11 and 12. The consistence of thermodynamic data with the presence of
a vHs near the Fermi level was shown in Ref. 13 by considering a p-d
like-model in the framework of slave-boson mean-field theory in the limit of
large U.

In this Letter we shall consider the two-dimensional single-band Hubbard
model by means of the Composite Operator Method (COM) [14,15]. In a standard
notation this model is described by the Hamiltonian 
\begin{eqnarray}
H=&&\sum_{ij} t_{ij} c^{\dagger}(i) \cdot c(j) +U\sum_i n_{\uparrow}(i)
n_{\downarrow}(i)  \nonumber \\
&&- \mu \sum_i c^{\dagger}(i)\cdot c(i)
\end{eqnarray}
To discuss the specific heat it is useful at first to consider the non
interacting [i.e. U=0] Hubbard model. In Fig. 1 we report $\gamma(x,T)$ as a
function of T for various values of the filling. We see that at half-filling 
$\gamma(x,T)$ diverges as $T\to 0$; this is an effect of the vHs. When
doping is introduced the Fermi energy moves away from the vHs and the peak
exhibited by $\gamma(x,T)$ moves away from $T=0$. $\gamma(x,T)$ firstly
increases as a function of T, exhibits a maximum at a certain temperature $%
T_m$ and then decreases. The temperature behavior of $\gamma(x,T)$ is
similar to the one exhibited by the static uniform spin magnetic
susceptibility $\chi_0(x,T)$ [16,5]. When we move away from half-filling the
value of $T_m$ increases up to a certain doping, then decreases. For a fixed
temperature $\gamma(x,T)$ exhibits a doping dependence of different
behaviour from the one presented by $\chi_0(x,T)$. At $T=0K$ $\gamma(x,T)$
diverges at $n=1$ and then decreases by lowering $n$. At nonzero temperature
the peak splits in two peaks, symmetric with respect to $n=1$ where it has a
minimum. This is shown in Fig. 2, where $\gamma(x,T)$ is reported as a
function of the filling at various temperatures. The shift of the two peaks
with respect to n=1 increases by increasing T.

When we consider the interacting case [i.e. $U\neq 0$], the specific heat in
COM is calculated by means of the following expression 
\begin{eqnarray}
C(T) &=&{\frac d{dT}}\int {\frac{d^2k}{(2\pi )^2}}\left\{ A_1\left( {\bf k}%
\right) f\left[ E_1\left( {\bf k}\right) \right] E_1\left( {\bf k}\right)
\right.   \nonumber \\
&&+\left. A_2\left( {\bf k}\right) f\left[ E_2\left( {\bf k}\right) \right]
E_2\left( {\bf k}\right) \right\} 
\end{eqnarray}
where $E_1({\bf k})$ and $E_2({\bf k})$ are the energy spectra; $A_1({\bf k})
$ and $A_2({\bf k})$ are some weight functions, calculated from the spectral
intensities of the Green's function, defined by $G({\bf k},\omega )=\langle
T[\psi (i)\psi ^{\dagger }(j)]\rangle _{F.T}$, where $\psi (i)$ is the
doublet composite operator 
\begin{equation}
\psi (i)=\left( 
\begin{array}{c}
\xi (i) \\ 
\eta (i)
\end{array}
\right) 
\end{equation}
with 
\begin{eqnarray}
\xi _\sigma (i) &=&c_\sigma (i)[1-n_{-\sigma }(i)] \\
\eta _\sigma (i) &=&c_\sigma (i)n_{-\sigma }(i)
\end{eqnarray}
By means of the equation of motion and by considering the static
approximation, where finite lifetime effects are neglected, the Green's
function $G({\bf k},\omega )$ can be computed in the course of a fully
self-consistent calculation where no adjustable parameters are considered
[14]. In Fig. 3 we present the linear coefficient $\gamma (x,T)=C(x,t)/T$ as
a function of the doping for various temperatures. As general behavior we
see that by increasing the doping $\gamma (x,T)$ increases up to a certain
doping and then decreases. The nature of the peak is again due to the fact
that the Fermi energy crosses the vHs for a certain critical value of $x$.
However, differently from the case of spin magnetic susceptibility, the peak
position of $\gamma (x,T)$ depends on the temperature. In the limit of zero
temperature a sharp peak is exactly located at $x=x_c$. By increasing the
temperature the peak moves away from $x_c$ and broadens in two peaks. The
situation is similar to what can be calculated for the free case and
illustrated in Figs. 1 and 2. The behavior described in Fig. 3 well
reproduces the experimental situation. As reported in Refs. 8 and 10, peaks
in the normal state linear coefficient are observed in $La_{2-x}Sr_xCuO_4$
and in $La_{2-x}Ba_xCuO_4$; the position of the peaks is close to the doping
where superconducting is suppressed, but there is a small shifting, due to
the temperature effect. In the case of $YBa_2Cu_3O_{6+y}$, the experimental
results reported in Ref. 9 are for the higher temperature $T=280K$; $\gamma
(x,T)$ increases with doping and presents two broad maxima in the region of
high doping. In Figs. 4 and 5 we present the linear coefficient $\gamma (x,T)
$ as a function of the temperature for values of the filling $x>x_c$ and $%
x<x_c$, respectively. At $x=x_c$ we see that $\gamma (x,T)$ diverges as $%
T\to 0$; this is an effect of the vHs. When $x\neq x_c$ the Fermi energy
moves away from the vHs and the peak exhibited by $\gamma (x,T)$ moves away
from $T=0$. $\gamma (x,T)$ firstly increases as a function of T, exhibits a
maximum at a certain temperature $T_m$ and then decreases. The behavior of $%
\gamma (x,T)$, when reported versus temperature, is similar to the one
exhibited by $\chi _0(T)$ [16,5]. As shown in Fig. 5, when the doping is
increased the value of $T_m$ moves to lower temperatures. This behavior
qualitatively agrees with the experimental situation reported in Ref. 9. The
fact that for $YBa_2Cu_3O_{6+y}$ $\gamma (x,T)$ is always a decreasing
function of T when $y>0.8$ indicates a low value of $T_m$, below the
critical superconducting temperature, which is the case near the critical
doping.

The main results obtained in this Letter can be so summarized. Experimental
data for the linear coefficient of the specific heat $\gamma$ in the normal
state of hole-doped cuprates show the existence of doping where the density
of states is enhanced, revealing the nearness of the Fermi level to the vHs.
This is in agreement with the ARPES experiments. The critical doping where
the Fermi level crosses the vHs is very close to the critical doping where
the superconducting phase is suppressed.

In the single-band Hubbard model the interaction has mainly two effects.
From one hand the critical doping is shifted from $x=0$ to some critical $%
x_c $. The value of $x_c$ depends on the ratio $U/t$ and varies between 0
and 1/3 [5]. For $U/t =4$ it is found that $x_c=0.27$, very close to the
experimental value observed in $La_{2-x}Sr_x CuO_4$ [16]. The shifting of
the Fermi energy with respect to the vHs by varying the doping explains and
well reproduces the unusual normal state behavior of $La_{2-x}Sr_x CuO_4$ in
the normal state of hole-doped cuprates. The other role played by the
interaction is a band structure effect which creates an asymmetric double
peak in the specific heat, as shown in Fig. 5. The picture of an ordinary
Fermi liquid which emerges from our calculations agrees with the
experimental situation in $YBa_2Cu_3O_{6+y}$ [17], where no separation of
spin and charge degrees of freedom has been observed. In the context of the
Hubbard model a van Hove scenario well describes some of the unusual
properties observed in the normal state; there is experimental evidence,
based on the data for susceptibility [16] and specific heat [8,10], that
this scenario is related to the overdoped region and not to the optimal
doping.

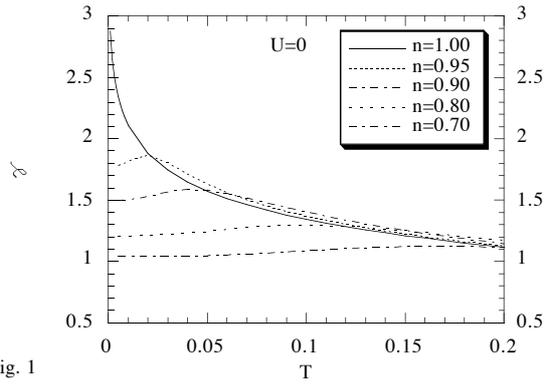
\begin{figure}[tbp]
\caption{The linear coefficient of the normal state specific heat $%
\gamma=C/T $ of the non interacting 2D Hubbard model is reported as a
function of the temperature for various values of the particle density. The
temperature is expressed in units of the hopping parameter $t$.}
\end{figure}

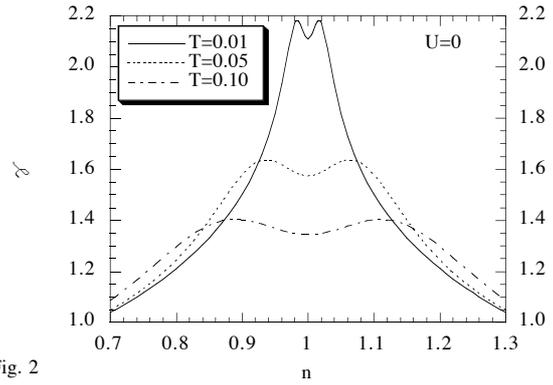
\begin{figure}[tbp]
\caption{$\gamma$ of the non interacting 2D Hubbard model is reported as a
function of the filling for various values of the temperature.}
\end{figure}

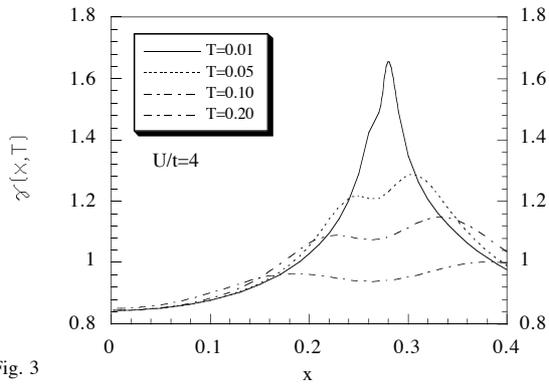
\begin{figure}[tbp]
\caption{$\gamma$ of the interacting 2D Hubbard model is reported as a
function of the doping parameter $x=1-n$ for various values of the
temperature and $U/t=4$.}
\end{figure}

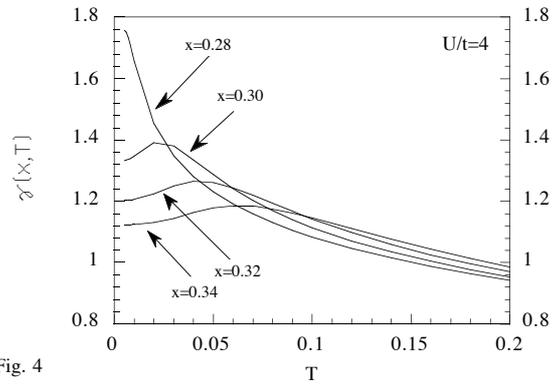
\begin{figure}[tbp]
\caption{$\gamma$ of the interacting 2D Hubbard model is reported as a
function of the temperature for various values of the doping $x>x_c$ and $%
U/t=4$.}
\end{figure}

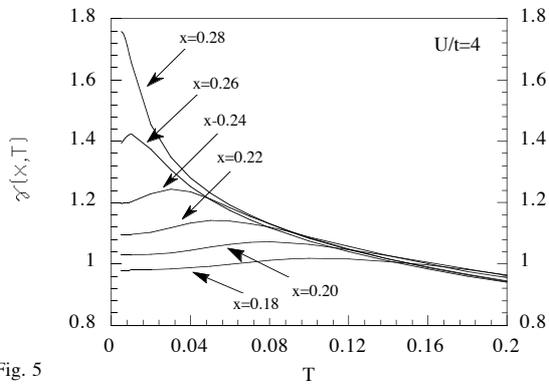
\begin{figure}[tbp]
\caption{Same as in Fig. 4, but for $x<x_c$.}
\end{figure}

\end{document}